\documentclass[twocolumn,aps,prb,groupedaddress,showpacs]{revtex4-1}
\usepackage{amsfonts}
\usepackage{amsmath}
\usepackage{amssymb}
\usepackage{txfonts}
\usepackage{pxfonts}
\usepackage{graphicx,bm,units,yfonts}
\usepackage[table]{xcolor}

\begin{document}

\title{Quantum criticality at the Chern-to-normal insulator transition}
\author{Yu Xue}
\author{Emil Prodan}
\email{prodan@yu.edu}
\affiliation{Department of Physics, Yeshiva University, New York, NY 10016, USA}
\date{\today }

\begin{abstract}
Using the noncommutative Kubo formula for aperiodic solids and a recently developed numerical implementation, we study the conductivity $\sigma$ and resistivity $\rho$ tensors as functions of Fermi level $E_F$ and temperature $T$ for models of strongly disordered Chern insulators. The formalism enabled us to converge the transport coefficients at temperatures low enough to enter the quantum critical regime at the Chern-to-trivial insulator transition. We find that the $\rho_{xx}$-curves at different temperatures intersect each other at one single critical point, and that they obey a single-parameter scaling law with an exponent close to the universally accepted value for the unitary symmetry class. However, when compared with the established experimental facts on the plateau-insulator transition in the Integer Quantum Hall Effect, we find a universal critical conductance $\sigma_{xx}^c$ twice as large, an ellipse rather than a semi-circle law, and absence of the Quantized Hall Insulator phase.
\end{abstract}

\pacs{72.25.-b, 72.10.Fk, 73.20.Jc, 73.43.-f}
\maketitle

\section{Introduction}

The criticality at the localization-delocalization transition (LDT) is believed to be universal and entirely determined by the generic symmetries of the systems (with certain special exceptions).\cite{EversRMP2008uy} One manifestation of this universality is a common scaling exponent $\nu$ per generic symmetry class, describing the diverging behavior of the localization length $\xi$ at the mobility edge:  $\xi \sim (E_F-E_F^c)^{-\nu}$. This in turn leads to a temperature-scaling of the transport coefficients at LDT:\cite{PruiskenPRL1988cm}
\begin{equation}\label{TScaling}
\rho(E_F, T) =F\left((E_F-E_F^c)(T/T_0)^{-\kappa}\right), \ \ \kappa= p/2\nu,
\end{equation}
when combined with the concept of temperature-induced effective size introduced by Thouless,\cite{ThoulessPRL1977fj} and with the single-parameter scaling hypothesis.\cite{Abrahams:1979et,AndersonPRB1980cx} In Eq. (\ref{TScaling}), $F$ is a system dependent function, $T_0$ is a reference temperature, and $p$ is the dynamical exponent for dissipation. Besides these universal scaling laws, there are other interesting universal aspects of the LDT, such as the existence of a single critical point (as opposed to a line of critical points) for the unitary symmetry class, universal values of the critical transport coefficients and universal renormalization flow-diagrams of the transport coefficients with the temperature or with the system-size.

Such universal characteristics at the plateau-plateau (PPT) and plateau-insulator (PIT) transitions in the Integer Quantum Hall Effect (IQHE) have preoccupied the experimental\cite{WeiPRB1985dh,KawajiJPSJ1987yu,WeiPRL1988jf,WakabayashiJPSJ1989bv,WakabayashiSurfSci1990df,KochPRB1991ls,DolgopolovJETP1991vc,KochPRL1991er,WeiPRB1992fh,WakabayashiJPSJ1992fj,KochSurfSci1992cv,AlphenaarPRB1994gh,ShaharPRL1995,ShaharSSC1998bn,HilkeNature1998fh,SchaijkPRL2000vc,ShimshoniMPLB2004rt,WanliPRL2005ty,VisserJPCS2006cu,WanliPRL2009gf} and theoretical\cite{WohlmanPRL1995gf,SondhiRMP1997sk,ShengPRB1999pp,PryadkoPRL1999uf,ShengPRB2000fr,ZulickeaPE2002bv,CainEPL2004bv,ShimshoniMPLB2004rt,DubiPRL2005tr,LevyArxiv2010bv,SongARXIV2013er} condensed matter communities for decades, resulting in some of the best experimental data and computer simulations available for a quantum transition. With the discovery of Topological Insulators (TI),\cite{HALDANE:1988rh,Kane:2005np,Kane:2005zw,BernevigScience2006hl,Koenig:2007ko,Moore:2007ew,Fu:2007vs,Hsieh:2008vm} the principles of universality will receive renewed scrutiny. The TIs have bulk extended states even in the presence of strong disorder,\cite{ProdanJPhysA2011xk} hence they are expected to display sharp LDTs. As such, the transport measurements at the transitions could be as clean and revealing as the ones in IQHE. The TIs can fall in different symmetry classes \cite{Schnyder:2008qy,Ryu:2010tq} and, even within the same symmetry class, the topological materials can be very different from one  another,\cite{ZHassanRevModPhys2010du,QiRMP2011tu} thus providing the perfect laboratory to test the principles of universality. The computer simulations have already began this process  (note that these are all zero-temperature finite-size scaling simulations).\cite{AsadaPRB2004bv,MarcosJPS2006ls,Onoda:2007xo,Obuse:2007qo,MedvedyevaPRB2010hg,FulgaPRB2011cv,FulgaPRB2012cg,NieuwenburgPRB2012fd,FulgaPRB2012jk,MongPRL2012ff,RossiPRL2012ww,YamakageArXiv2012jf} One question that received great attention from these works is if the scaling exponents of the symplectic models at the metal-to-normal insulator and at the metal-to-topological insulator are the same. So far, the answer seems to be affirmative.

Although there is a substantial experimental transport data for TIs, only recently the focused was fully tuned on the topological-to-nontopological phase transition.\cite{SatoNatPhys2011hf,XuScience2011gg,BrahlekPRL2012hh} And even these experiments need to be further refined for the quantum criticality to be revealed. Traditionally, the experiment was always ahead of the theoretical simulations in this domain (see Ref.~\onlinecite{SongARXIV2013er} for a discussion), but we strongly believe that this state of affairs will be soon reversed by the adoption of the non-commutative geometry program for aperiodic solids, initiated by Bellissard {\em et al} in the 90's.\cite{BELLISSARD:1994xj,Schulz-Baldes:1998oq,Schulz-Baldes:1998vm} For example, this natural formalism for treating disordered solids under magnetic fields enabled us to develop extremely accurate, stable and efficient simulations of the zero- and finite-temperature transport coefficients,\cite{Prodan2010ew,ProdanJPhysA2011xk,ProdanAMRX2012bn,Xue2012fh} to a point where qualitative and quantitative agreement between experiment and simulation was possible for PIT in IQHE.\cite{SongARXIV2013er} In this article, we announce several predictions based on the noncommutative Kubo formula on the quantum criticality at the Chern-to-normal insulator transition. On a broader note, we want to announce that this type of simulations reached a level where they provide qualitative and quantitative guidance for the experiments on quantum criticality at the topological-to-nontopological phase transitions. 

The search for possible Chern Insulator (CI) materials have intensified and several theoretical studies have already singled out possible CI candidates.\cite{WuPRL2008xx,InouePRL2010bv,XuPRL2011gh,ZhangPRL2011bb,XuPRB2012vu,EssinPRB2012bv,ZhangPRB2012bv,RueggPRB2012gg,LiPRB2012nn,EzawaPRL2012jj,EzawaPRL2013er,ZhangArXiv2013hf} This and the fact that the CIs and the IQHE liquids belong to the same unitary symmetry class motivated us to focus exclusively on CIs in this study. For the spin-up sector of the Kane-Mele (KM) model\cite{Kane:2005np} with strong disorder, we were able to converge ({\it i.e.} eliminate any finite-size effects) the finite-temperature conductivity $\sigma$ and resistivity $\rho=\sigma^{-1}$ tensors, at temperatures low enough to enter the quantum critical regime at the transition between a CI and a normal insulator. We compare these results with the known facts for the PIT in IQHE.\cite{AlphenaarPRB1994gh,ShaharSSC1998bn,HilkeNature1998fh,SchaijkPRL2000vc,DunfordaPE2000gf,PonomarenkoPE2000yt,ShimshoniMPLB2004rt,PruiskenSSC2006tr,VisserJPCS2006cu,LangPRB2007yg} 

We find that, like at PIT in IQHE, the graphs of $\rho_{xx}$ as function of electron density, recorded at different temperatures, intersect at one single critical point, and they collapse into a single curve after a single-parameter rescaling. The scaling exponent $\kappa$ is in good agreement with what one would predict by using the universally accepted value of the finite-size scaling exponent $\nu = 2.58\pm 0.03$,\cite{SlevinPRB2009tr,KramerIJMPB2010wk,ObusePRB2010fj,FulgaPRB2011cv,DahlhausPRB2011bn,AmadoPRL2011fj,SlevinIJMP2012gh} and with $p$ fixed like in our simulations ($p=1$). We clearly see the expected renormalization flow of $\sigma$ with the temperature, but the separatix is not a semicircle like at PIT in IQHE, but rather an ellipse. At the critical point, we find $\sigma_{xy} \approx \frac{1}{2} \frac{e^2}{h}$ and, quite interestingly, $\sigma_{xx} \approx \frac{e^2}{h}$ rather than $\frac{1}{2} \frac{e^2}{h}$ (to be precise, the numerics place $\sigma_{xy}$ between $0.5-0.6 \frac{e^2}{h}$ at PIT \cite{HuoPRL1993bf,GammelPRL1994tr,WangPRL1996tt,ChoPRB1997vc,SchweitzerPRL2005bv,MarcosJPS2006ls}). The main surprise was, however, the absence of the Quantized Hall Insulator phase, characterized by $\sigma=0$ but $\rho_{xy}=\frac{h}{e^2}$.

To probe the broader applicability of our conclusions, we have repeated the computations for the spin-up sector of the Bernevig-Hughes-Zhang (BHZ) model.\cite{BernevigScience2006hl} Although in this case the calculations are not converged well enough for accurate quantitative predictions, they already support  the qualitative conclusions we reached for the KM model: existence of a single critical point and absence of the Quantized Hall Insulator pahse. The critical values of the transport coefficients remain the same. Unfortunately, the data for BHZ model display a poorer scaling, indicating that lower temperatures are needed. As such, the scaling exponent for this second set of calculations is inconclusive. We plan to re-examine the issue in future simulations with larger system-sizes, which will enable us to further reduce the temperature without introducing size effects. 

\section{The noncommutative Kubo formula}

The key for our simulations is the noncommutative Kubo formula:\cite{BELLISSARD:1994xj,Schulz-Baldes:1998oq,Schulz-Baldes:1998vm}
\begin{equation}\label{Kubo}
\sigma_{ij}(E_F,T)=-{\cal T}\left([x_i,H](1/\tau+{\cal L}_H)^{-1}[x_j,\Phi_{\mathrm{FD}}(H)]\right),
\end{equation} 
where ${\cal T}$ represents the trace over volume, $H$ is the disordered Hamiltonian, ${\bm x}$ is the position operator, $\tau$ is the relaxation time for dissipation, $\Phi_{\mathrm{FD}}$ is the Fermi-Dirac distribution for a given $T$ and $E_F$, and ${\cal L}_H$ is the Liouvillian acting on the observables (i.e., operators):
\begin{equation}
{\cal L}_H A= i[H,A].
\end{equation}
 In real condensed matter systems: $\tau \sim T^{-p}$, where $p$ is the dynamical exponent appearing in Eq.~(\ref{TScaling}). 

As usual, the conductivity tensor $\sigma_{jk}$ of Eq.~(\ref{Kubo}) provides the link between the charge current density and the homogeneous static electric field ${\bm E}$ that drives the system:
\begin{equation}
J_j = \sum_{k=1}^d \sigma_{jk} E_k.
\end{equation}
Here, ${\bm J}$ is the time average of the expected value of the current density operator. The state of the system evolves according to the dynamics induced by $H-e{\bm E}{\bm x}$, and by random (in time), instantaneous electron scattering events. Realistic electron-electron and electron-phonon scattering matrices can be considered, however, Eq.~(\ref{Kubo}) assumes the relaxation time approximation.

Extended discussions of the physical assumptions, of the derivation and of the noncommutative formalism, together with convergence tests and applications to the disordered Hofstadter model can be found in Ref.~\onlinecite{ProdanAMRX2012bn}. Another complete discussion of the Kubo formalism was given in this very recent journal (see Ref.~\onlinecite{Xue2012fh}), together with applications to a disordered model of a quantum spin-Hall insulator. For these reasons, here we only discuss the significance of Eq.~(\ref{Kubo}) and its relation with other works. Eq.~(\ref{Kubo}) represents the thermodynamic limit of the formal Kubo formulas written in terms of the eigenfunctions and eigenvalues of the equilibrium Hamiltonian, found in the classical solid state textbooks [see for example Eq.~(3.385) and its finite-temperature version in Ref.~\onlinecite{MahanBook2000bv}]. Finite-volume versions of the Kubo formula for aperiodic systems can be also derived using the traditional quantum master equation.\cite{GebauerLNP2006bf} In this type of analyses, the Kubo formula is formulated as the infinite-volume limit of the finite-volume expressions. The existence of this limit for aperiodic systems is virtually impossible to prove without the noncommutative formalism. What is special about the latter is that it enables one to work directly in the thermodynamic limit. For example, the trace per volume is rewritten using the Birkhoff ergodic theorem, and the Liouvillian can be defined as a normal operator on a well defined Hilbert space (hence a spectral decomposition exists for it). 

At the practical level, an explicit Kubo formula for infinite volume is important because it enables one to analyze how fast are the various finite-volume approximations converging in the thermodynamic limit. In particular, it enabled us to develop a canonical finite-volume approximation that converges exponentially fast in the thermodynamic limit, a fact that was established with mathematical rigor.\cite{ProdanAMRX2012bn} This canonical finite-volume approximation and its numerical implementation have been extensively discussed in Refs.~\onlinecite{ProdanAMRX2012bn,Xue2012fh}. Here we closely follow these two references.

\section{Quantum criticality in the spin-up sector of the Kane-Mele model}

 \begin{figure}
\center
  \includegraphics[height=5cm]{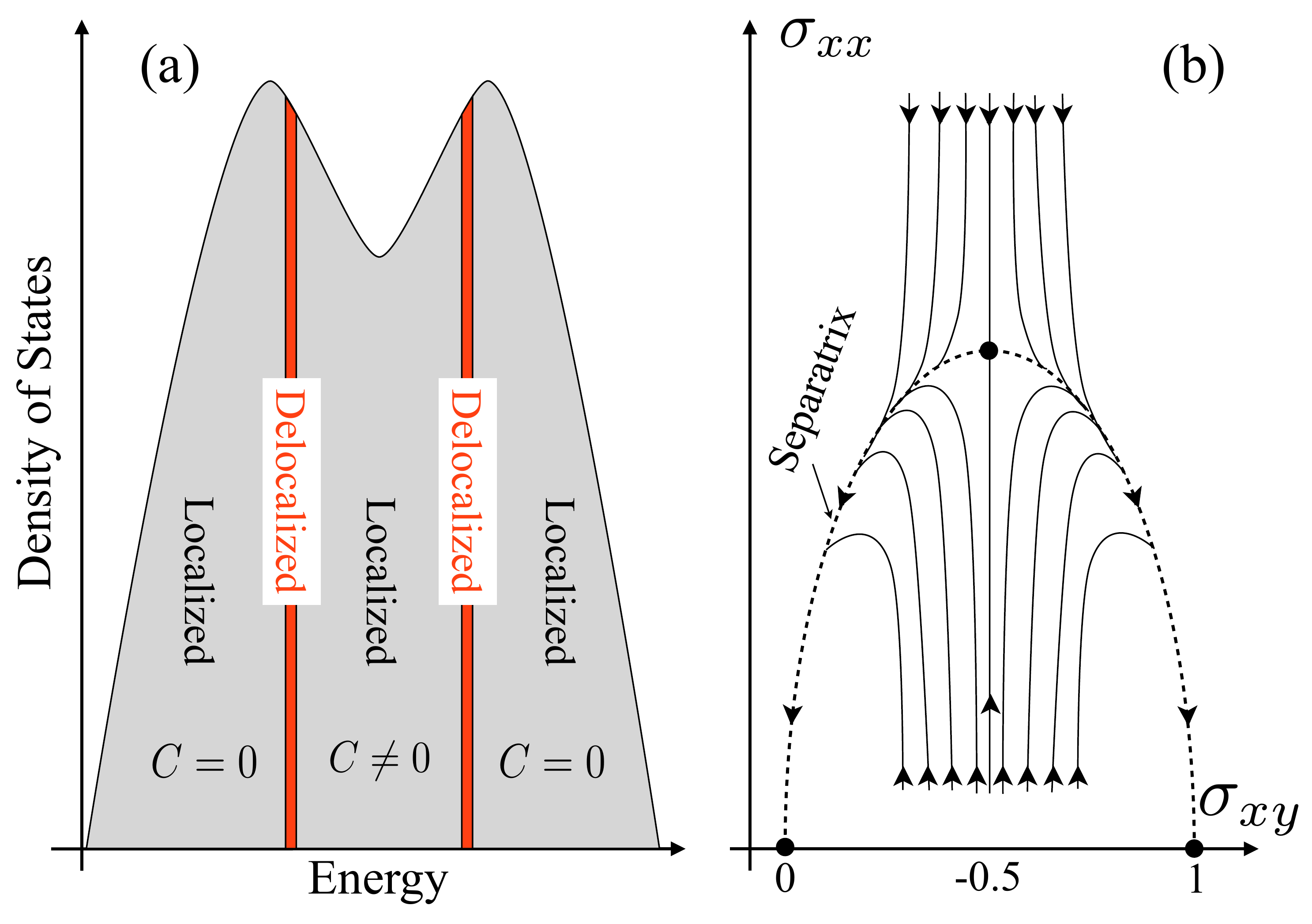}\\
  \caption{(Color online) (a) The spectrum of the disordered Chern insulator is localized everywhere except for two thin energy regions. (b) The expected flow of the transport coefficients when the temperature is lowered to zero. Different flow lines correspond to different Fermi levels. The dashed line represents the separatrix, which is expected to satisfy a semicircle law.}
 \label{Summary}
\end{figure}

In the absence of Rashba interaction, the spin sectors decouples and the KM model reduces to a disconnected sum of two Chern insulators. The Hamiltonian in the spin-up sector (tuned in the middle of the CI topological phase) takes the form: 
\begin{equation}\label{KM}
H_0=\sum\limits_{\langle {\bm n \bm m} \rangle} |{\bm n}\rangle \langle {\bm m}| +0.6 \mathrm{i} \sum\limits_{\langle \langle {\bm n \bm m} \rangle \rangle}  \tau_{\bm n}\big ( |{\bm  n}\rangle \langle {\bm  m}|- |{\bm  m}\rangle \langle {\bm  n}|\big ).
\end{equation}
Here, $\tau_{\bm n}$ represents the isospin of the site ${\bm n}$ of the honeycomb lattice, and $\langle  \rangle$/$\langle \langle \rangle \rangle$ symbolize first/second nearest neighbors. We add  the random potential $V_\omega=W\sum_{{\bm n}}\omega_{{\bm n}} |{\bm n}\rangle \langle {\bm n}|$ to $H_0$, where the $\omega$'s are independent random variables uniformly distributed in $[-\frac{1}{2},\frac{1}{2}]$. We fix $W=4$ ($=2 \ \times$ the clean insulating gap), in order to achieve the strong disorder regime where the insulating (spectral) gap is closed and only a mobility gap remains. The simulations are performed on a finite-size lattice containing $80 \times 80$ unit cells. By repeating the simulations with different lattice sizes, we concluded that the effects due to the finite-size of the simulation box are practically negligible for the temperatures considered in the present study (the effective Thouless length\cite{ThoulessPRL1977fj} is smaller than the simulation box).   

\begin{figure}
\center
  \includegraphics[height=7cm]{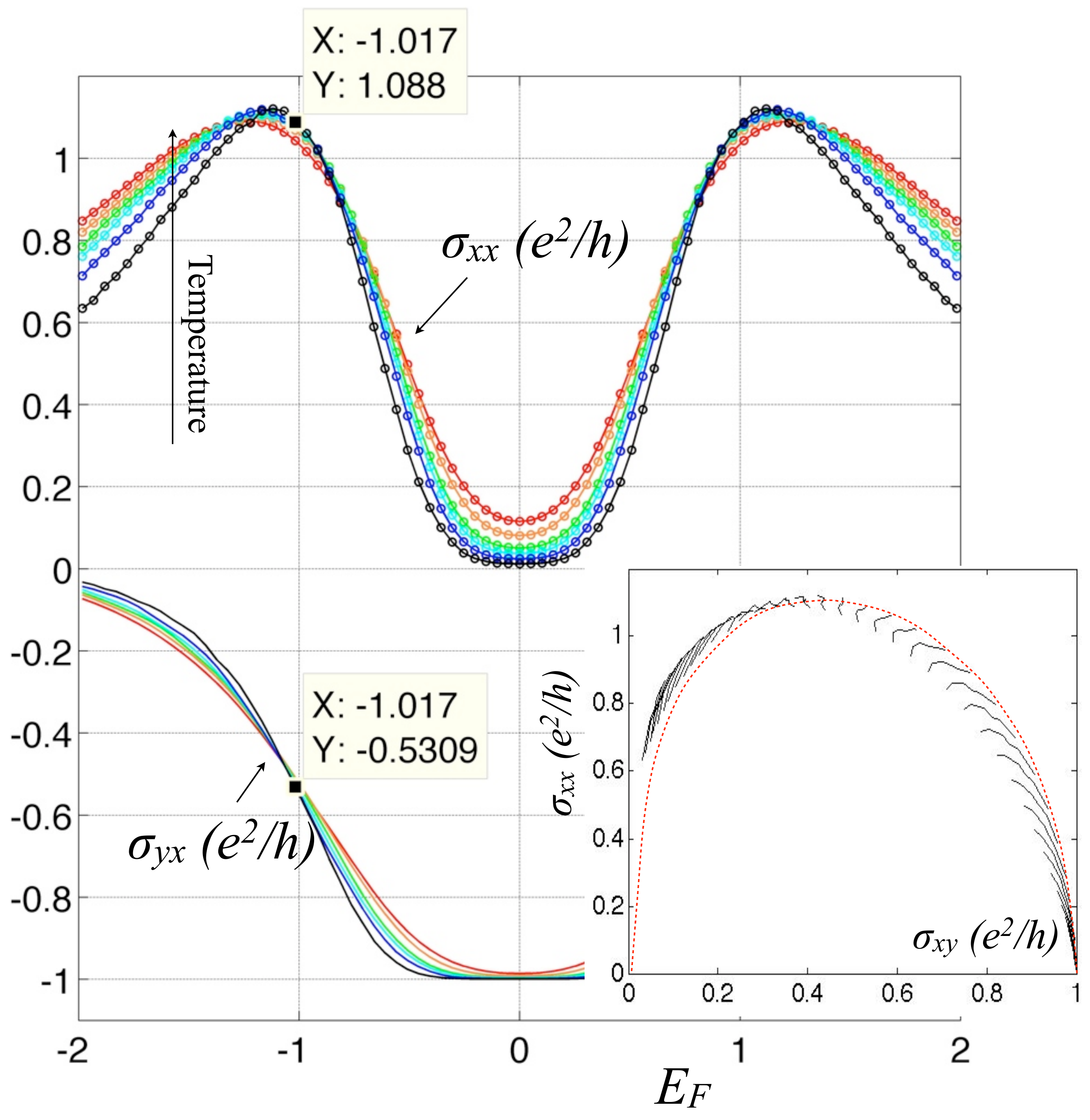}\\
  \caption{(Color online) $\sigma$ as function of $E_F$ at $kT=1/\tau=0.01$, 0.02, 0.03, 0.04, 0.06, 0.08, simulated on an $80 \times 80$ lattice. An average over many disorder configurations was considered (as many as 67 for $T=0.01$ and 24 for $T=0.08$). The marks posted on the graph give $\sigma$ at the critical point. The inset shows the flow of our data in the $(\sigma_{xy},\sigma_{xx})$ plane as $T \rightarrow 0$. The dotted line suggests the separatrix for the flow.}
 \label{KMSigmaVsEf}
\end{figure} 

In the above conditions, all the quantum states of the model are localized, except for the states in two narrow energy regions separating the CI from the normal insulator [see Fig.~\ref{Summary}(a)]. The existence of such delocalized quantum states can be demonstrated\cite{ProdanJPhysA2011xk} with mathematical rigor using the theory of noncommutative Chern number,\cite{BELLISSARD:1994xj} while numerically, it has been demonstrated using recursive Green's function and transfer matrix calculations,\cite{Onoda:2007xo,Yamakage2010xr,XuPRB2012vu,SongPRB2012gf} level statistics analysis,\cite{Prodan2010ew,ProdanJPhysA2011xk,ChuaPRB2011ci} simulations of the edge currents and computations of the edge conductance.\cite{LiPRL2009xi,GrothPRL2009xi,JiangPRB2009sf} Near the transitions, the field-theoretic arguments developed by Pruisken and collaborators \cite{PruiskenNP1984fj,LevineNP1984tt,LevineNP1884hd,LevineNP1984uu} for IQHE predict the $T$-driven flow-diagram shown in Fig.~\ref{Summary}(b), which was observed and confirmed in IQHE by experiment.\cite{KawajiJPSJ1987yu,YamaneJPSJ1989bv}  

\begin{figure}
\center
  \includegraphics[height=6.5cm]{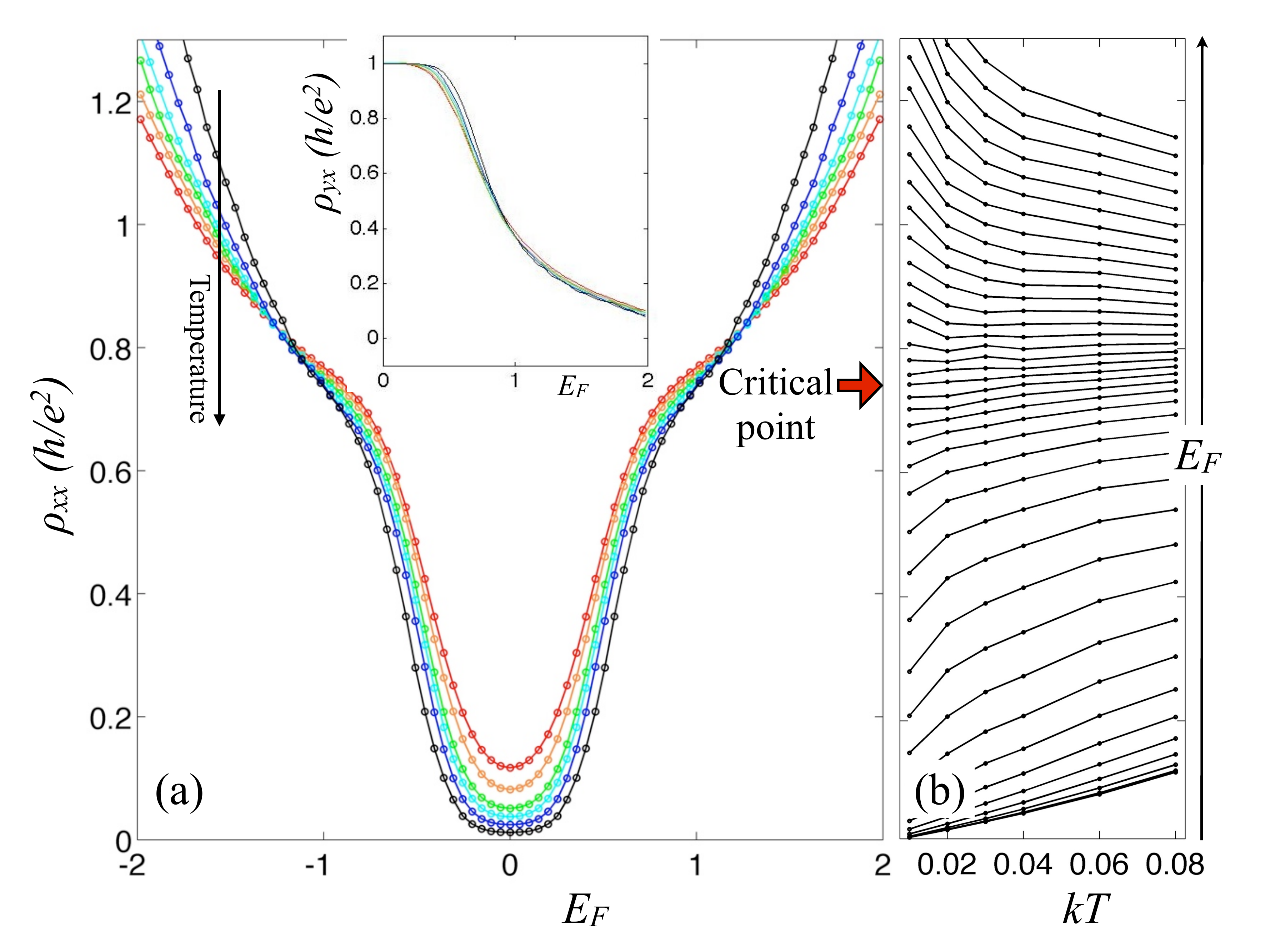}
  \caption{(Color online) (a) $\rho_{xx}$ (and $\rho_{xy}$ in the inset) as function of $E_F$ at $kT=0.01$, 0.02, 0.03, 0.04, 0.06 and 0.08. (b) $\rho_{xx}$ as function of temperature for various $E_F$ values. The arrow indicates the transition from CI to the normal insulator.}
 \label{KMRhoVsEf}
\end{figure}

The simulated $\sigma$ is reported in Fig.~\ref{KMSigmaVsEf} as function of Fermi level, for $kT=1/\tau=0.01$, $0.02$, $0.03$, $0.04$, $0.06$ and $0.08$ (hence we fix $p=1$ in our simulations). In this figure we can see an energy region where, especially for the lower temperatures, $\sigma_{yx}$ takes the quantized value of $-1$, indicating that the system is in the CI phase. When moving away from this energy region, $\sigma_{yx}$ starts to converge towards $0$, indicating that the system enters the normal insulator phase. The $\sigma_{yx}$ curves computed at different temperatures intersect each other at practically one point. Examining the data for the direct conductivity, we see $\sigma_{xx}$ decreasing as $T \rightarrow 0$ for most part of the energy spectrum, a hallmark of the insulating phase, with the exception of two distinct energy regions where $\sigma_{xx}$ increases as temperature decreases. These energy regions appear exactly where $\sigma_{yx}$ switches between its quantized values and, as such, they must harbor extended quantum states.\cite{BELLISSARD:1994xj} The energy regions where $\sigma_{xx}$ increases as $T \rightarrow 0$ appear to become smaller and smaller as $T$ is lowered, and that the maximum value of $\sigma_{xx}$ stabilizes at a finite value (as opposed to diverging to infinity). An important question is if these regions reduce to a point as $T \rightarrow 0$. A more refined analysis based on Fig.~\ref{KMRhoVsEf} shows that this is, indeed, the case, and gives the critical Fermi levels $E_F^c\approx \pm 1.02$. The values of $\sigma$ at $E_F^c$ are marked in Fig.~\ref{KMSigmaVsEf} and are $\sigma_{xx}^c \approx \frac{e^2}{h}$ and $\sigma_{xy}^c \approx \frac{1}{2}\frac{e^2}{h}$. These critical values for the conductance are reproduced when the simulations are repeated for the spin-up sector of the BHZ model. Furthermore, the PIT is known to exhibit a universal critical conductance,\cite{ShaharPRL1995} though with a different value, so we are led to conjecture that the Chern-to-normal insulator transition also exhibits a universal conductance, with the universal values stated above. 

When the data points from the main plot in Fig.~\ref{KMSigmaVsEf} are placed in the $(\sigma_{xy},\sigma_{xx})$ plane, we obtain the flow-diagram shown in the inset of Fig.~\ref{KMSigmaVsEf}. The separatrix for this flow, marked with the dotted line, has the shape of a semiellipse with the semiaxes $\frac{1}{2}\frac{e^2}{h}$ and $\frac{e^2}{h}$. At PIT in the IQHE, the separatrix strictly obeys the semicircle law: $\sigma_{xx}^2+(\sigma_{xy} - 0.5 e^2/h)^2=(0.5 e^2/h)^2$.\cite{ShaharSSC1998bn} Let us point out that the values $\sigma_{xy}>0.5$ occur when $E_F$ is located inside the old clean insulating gap, while the values $\sigma_{xy}<0.5$ occur when $E_F$ is located in the high-density spectrum resulted from the localization of the clean energy bands. As such, the flow (as $T \rightarrow 0$) starts from the inside (outside) of the ellipse and moves towards the separatrix for $\sigma_{xy}>0.5$ ($\sigma_{xy}<0.5$), a markedly different behavior when compared with PIT.

 \begin{figure}
\center
  \includegraphics[height=6.5cm]{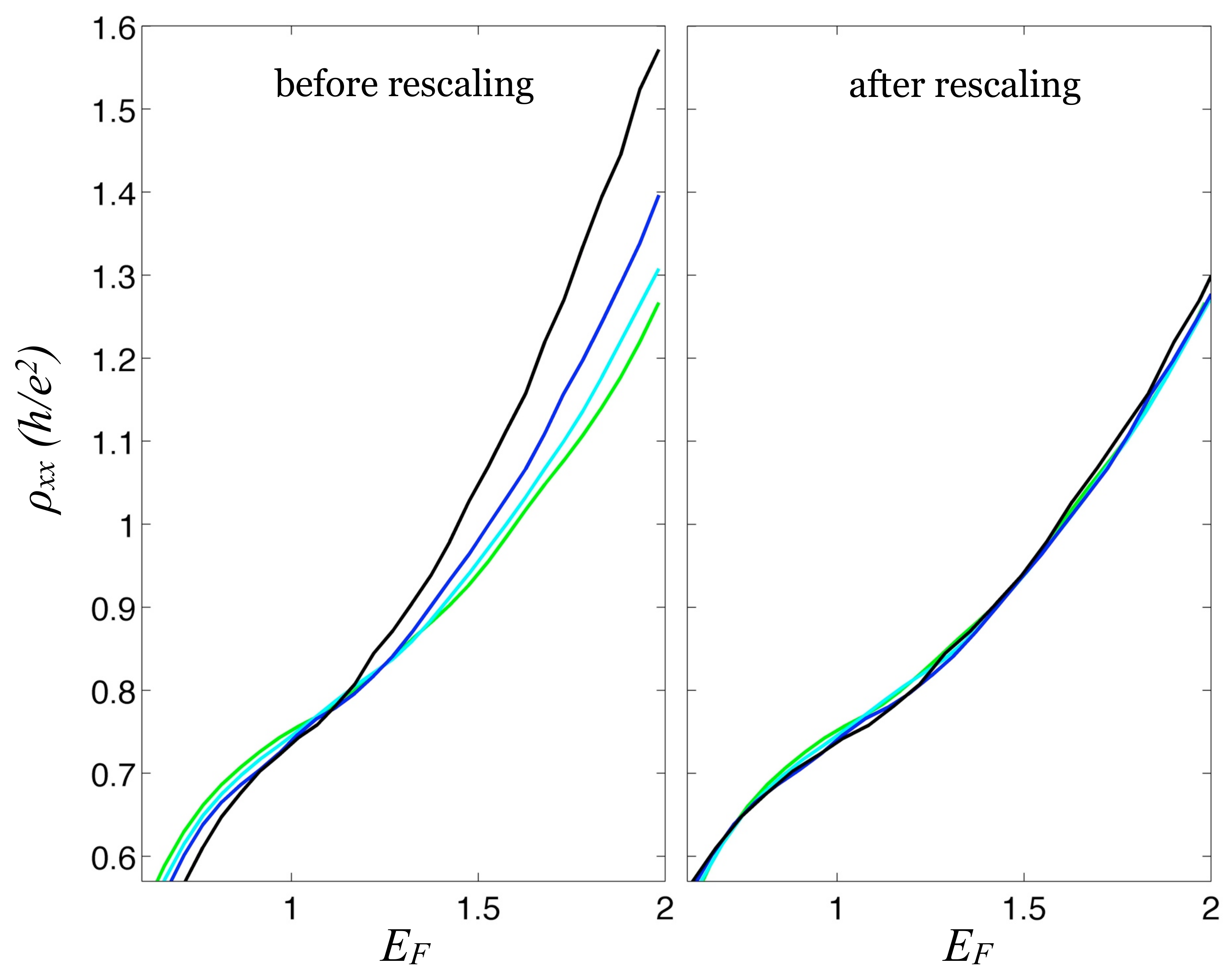}
  \caption{(Color online) $\rho_{xx}$ at different temperatures collapses onto a single curve after the single-parameter rescaling: $E_F \rightarrow E_F^c+(E_F-E_F^c)(T/T_0)^{-\kappa}$, with $E_F^c=1.017$, $T_0=0.04$ and $\kappa=0.21$.}
 \label{KMRescaling}
\end{figure}

The existence of a unique critical point (as opposed to a line of critical points) at the Chern-to-normal transition can be established with great confidence by examining the resistivity tensor, plotted in Fig.~\ref{KMRhoVsEf}. From the expression $\rho_{xx}=\sigma_{xx}/(\sigma_{xx}^2+\sigma_{xy}^2)$, it follows that $\rho_{xx} \rightarrow 0$ inside the CI phase, while $\rho_{xx} \rightarrow \infty$ inside the normal insulating phase, when $T\rightarrow 0$. As a consequence, the $\rho_{xx}$ curves at different $T$ must cross each other, very much like the $\sigma_{yx}$ curves do in Fig.~\ref{KMSigmaVsEf}. Since the plots are symmetric relative to the zero energy, we can focus only on the positive energies. Fig.~\ref{KMRhoVsEf}(a) strongly suggests that all $\rho_{xx}$ curves cross each other at a single point, exactly how it was observed at PIT. To accurately pin-point this crossing point, we replotted $\rho_{xx}$ in Fig.~\ref{KMRhoVsEf}(b), this time as function of temperature, for each positive $E_F$-value appearing in Fig.~\ref{KMRhoVsEf}(a). The flow of $\rho_{xx}$ with $T$ curves down at lower $E_F$'s and it curves up at higher $E_F$'s. There is a clear separatrix between these two distinct tendencies, very much like in the experimental data on the PI transition in Ref.~\onlinecite{MurphyPE2000yy}, or on the  metal-insulator transition in Ref.~\onlinecite{KnyazevPRL2008ui}. This enables us to accurately identify the critical point: $E_F^c = \pm 1.017$, and then to determine the value of the conductivity tensor at the critical point: $\sigma_{xy}^c=-0.53\times \frac{e^2}{h}$ and $\sigma_{xx}^c= 1.09 \times \frac{e^2}{h}$. 

The inset in Fig.~\ref{KMRhoVsEf}(a) reports $\rho_{yx}=\sigma_{xy}/(\sigma_{xx}^2+\sigma_{xy}^2)$ as function of $E_F$, which decreases from 1 to 0 almost with the same rate as $\sigma_{yx}$. As one can see, there is absolutely no tendency for $\rho_{xy}$ to stay quantized at $\frac{h}{e^2}$ through the transition or further into the normal insulating phase. Such quantization of $\rho_{xy}$ would have been incompatible with the critical values of $\sigma$ determined above. It is also known that a quantized $\rho_{xy}$ is equivalent with the semicircle law,\cite{HilkeNature1998fh} but the separatrix shown in the inset of Fig.~\ref{KMSigmaVsEf} has an elliptical shape quite different from a semicircle. We want to point out that these facts are also true for the simulations with the BHZ model.  As such, we can conclude with great confidence that the Quantized Hall Insulator phase is absent for this system. This is in striking contrast with the PIT in IQHE, for which we did observed the Quantized Hall Insulating phase,\cite{SongARXIV2013er} using same type of calculations. 

We now zoom into the region around $E_F^c$ and start the scaling analysis. Since the scaling occur in the asymptotic limit $T \rightarrow 0$, we keep for this analysis only the lowest four temperatures. As shown in Fig.~\ref{KMRescaling}, the $\rho_{xx}$-curves collapse almost perfectly on top of each other after the energy axis is rescaled as: $E_F \rightarrow E_F^c+(E_F-E_F^c)(kT/kT_0)^{-\kappa}$ ($kT_0=0.04$). The best overlap of the rescaled curves is obtained for $\kappa=0.21 \pm 0.01$, a value that is in good agreement with $k=0.194 \pm 0.002$ obtained from the expression $\kappa=p/(2\nu)$ with the universally accepted value $\nu=2.58 \pm 0.03$, and $p=1$ like in our simulations.

\section{Critical regime in the spin-up sector of the Bernevig-Hughes-Zhang model}

\begin{figure}
  \includegraphics[height=6.5cm]{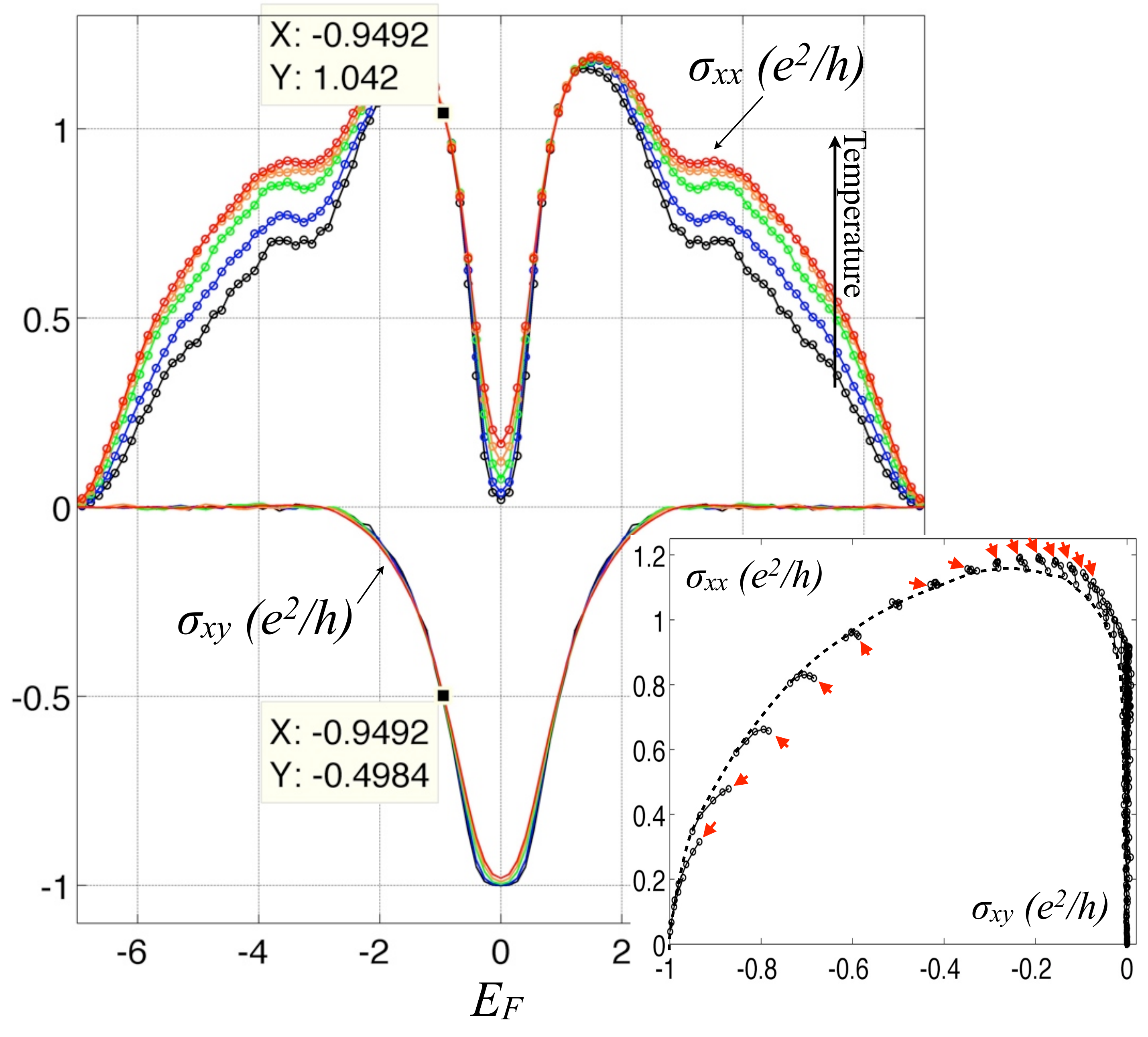}\\
  \caption{(Color online) The simulated $\sigma$ for BHZ model as function of $E_F$ at $kT=1/\tau=0.01$, 0.02, 0.04, 0.06, 0.08, simulated on an $80 \times 80$ lattice. The inset shows the $T$-driven renormalization flow of $\sigma$ in the $(\sigma_{xy},\sigma_{xx})$ plane.}
 \label{BHZSigmaVsEf}
\end{figure}

This section reports the simulations based on the non-commutative Kubo formula, for the Chern Insulator (CI) corresponding to the spin-up sector of the Bernevig-Hughes-Zhang (BHZ) model (tuned in the middle of the topological phase):\cite{BernevigScience2006hl} 
\begin{equation}\label{BHZ}
h({\bm k})=\sigma_x \sin k_x+ \sigma_y \sin k_y+2(1+\cos k_x + \cos k_y)\sigma_z,
\end{equation}
where the $\sigma's$ represent the Pauli's matrices. This model can be represented on a 2-dimensional square-lattice with two orbitals per site. The lattice sites are indexed by ${\bm n}$ and the orbitals by $\alpha$. In this real-space representation, we add the random potential $V_\omega=W\sum_{{\bm n},\alpha}\omega_{{\bm n},\alpha} c^\dagger_{{\bm n},\alpha}c_{{\bm n},\alpha}$, where $c^\dagger_{{\bm n},\alpha}$ creates an electron in state $\alpha$ at site ${\bm n}$, and $\omega$'s are independent random variables uniformly distributed in $[-1/2,1/2]$. We fixed $W=5$ ($=2.5 \ \times$ the clean insulating gap), to achieve the strong disorder regime where the insulating gap is closed and only a mobility gap remains. As in the previous section, the lattice-size was taken to be $80 \times 80$ unit cells. An average over many disorder configurations was considered for different temperatures, specifically: $68$ configurations for $kT=0.01$, $67$ for $kT=0.02$ and $23$ for $kT=0.04$, $0.06$ and $0.08$. 

\begin{figure}
  \includegraphics[height=6cm]{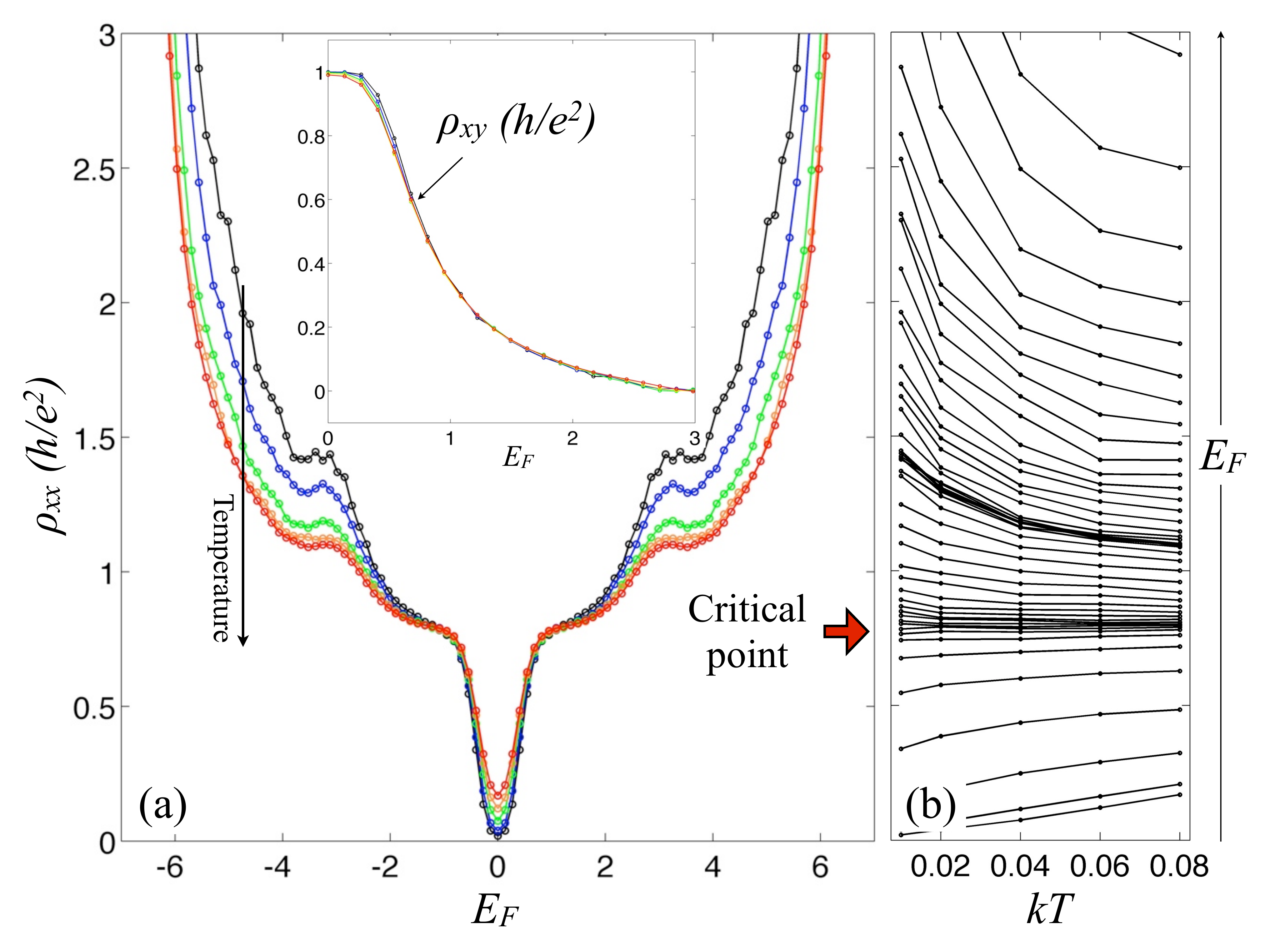}
  \caption{(Color online) (a) $\rho_{xx}$ for BHZ model as function of $E_F$ at $kT=0.01$, 0.02, 0.04, 0.06 and 0.08. The inset reports $\rho_{xy}$. (b) $\rho_{xx}$ as function of temperature for various $E_F$ values. The arrow indicates the transition from CI to the normal insulator.}
 \label{BHZRhoVsEf}
\end{figure}  

Fig.~\ref{BHZSigmaVsEf} reports the simulated conductivity tensor $\sigma$ as function of Fermi level $E_F$, at various temperatures $T$. The data display the same general features as seen in Fig.~\ref{KMSigmaVsEf}. In particular, $\sigma_{xy}$ takes quantized values in certain energy regions and, in the energy region where $\sigma_{xy}$ shifts between the quantized values, $\sigma_{xx}$ becomes independent of temperature, signaling the presence of extended states. A more refined analysis based on Fig.~\ref{BHZRhoVsEf} reveals again a single critical point, located at the critical Fermi level $E_F^c\approx 0.94$. The critical $\sigma$ values at $E_F^c$ were found again to be $\sigma_{xx}^c \approx \frac{e^2}{h}$ and $\sigma_{xy}^c \approx \frac{1}{2}\frac{e^2}{h}$ (the exact values are marked on the graph). The inset of Fig.~\ref{BHZSigmaVsEf} shows the $T$-driven flow of $\sigma$ in the $(\sigma_{xy},\sigma_{xx})$ plane. While it generally resembles the flow in Fig.~\ref{KMSigmaVsEf}, a closer analysis reveals that the calculations are not well converged on the normal insulator side (near $\sigma_{xy}=0$). Lower temperatures will be definitely needed to obtain the correct separatrix (which we believe it will look very similar to that for KM model).

Fig.~\ref{BHZRhoVsEf} reports the simulated resistivity tensor $\rho$ (a) as function of $E_F$, at various temperatures, and (b) as function of $kT$, at various Fermi levels. The flow with $T$ in Fig.~\ref{BHZRhoVsEf}(b) is used to accurately determined the critical point, whose coordinates were given above. Apart from some structure in the curves, there are no major differences when Fig.~\ref{BHZRhoVsEf} is compared with Fig.~\ref{KMRhoVsEf}. The inset reports $\rho_{xy}$ as function of $E_F$ and, here too, there is no trace of the Quantized Hall Insulator phase.

\begin{figure}
  \includegraphics[height=5cm]{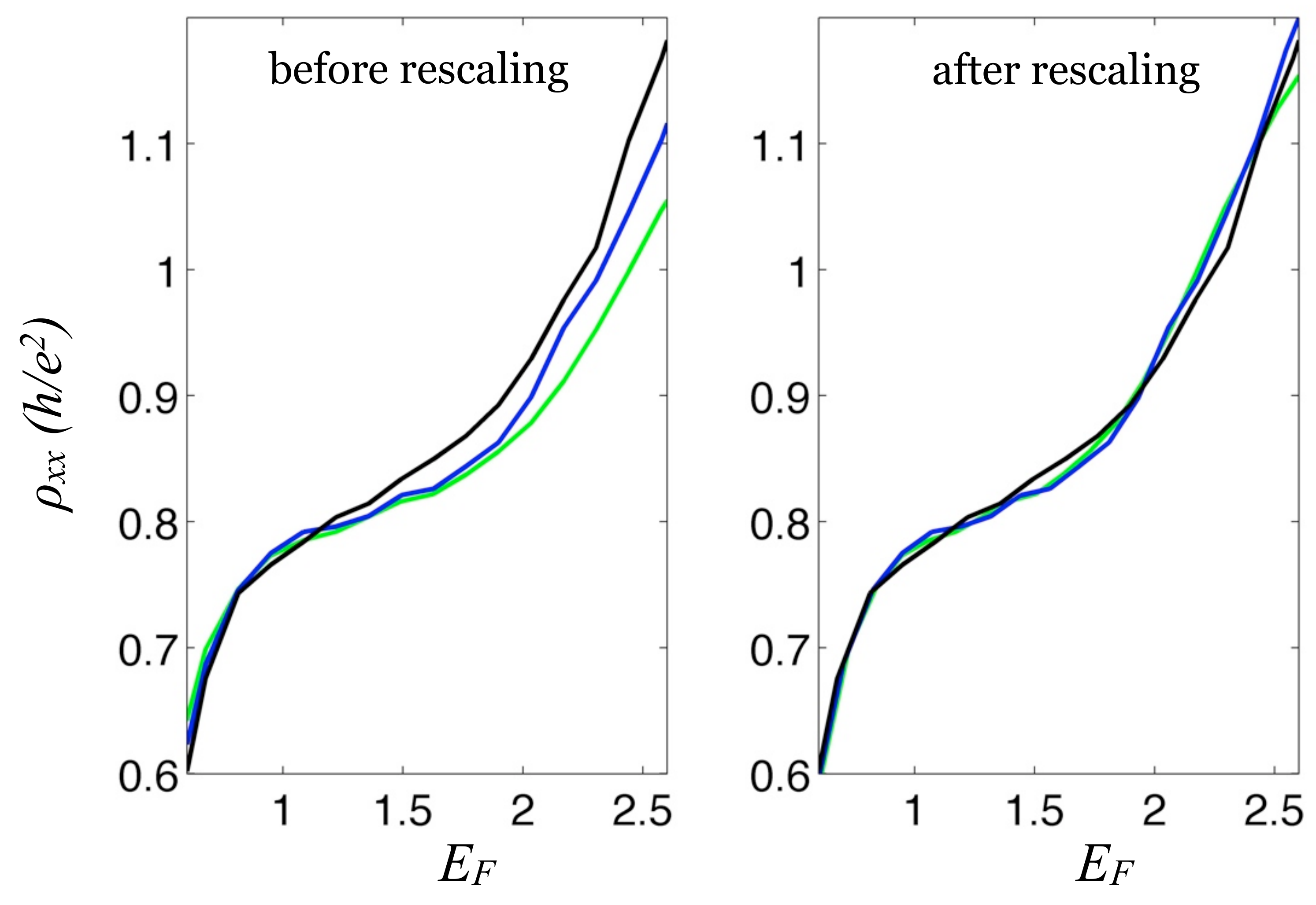}
  \caption{(Color online) Single-parameter rescaling for BHZ model: $E_F \rightarrow E_F^c+(E_F-E_F^c)(T/T_0)^{-\kappa}$, with $E_F^c=0.94$, $T_0=0.01$ and $\kappa=0.14$.}
 \label{BHZRescaling}
\end{figure}

Fig.~\ref{BHZRescaling} reports the scaling analysis for the BHZ model. Only the lowest three temperatures have been considered. The best overlap of the rescaled curves is obtained for $\kappa=0.14 \pm 0.01$, but as already anticipated, the overlap is quite poor. As such, we conclude that this numerical value is inconclusive. 

When searching for a reason for the poorer convergence of the results in the BHZ model, we found that, in all our previous simulations,\cite{Prodan2010ew,ProdanJPhysA2011xk,Prodan2011vy,XuPRB2012vu} the band energy states in the BHZ model localize much slower than in the KM model. The only major qualitative difference between KM and BHZ models is that the latter has one split-Dirac point, located at ${\bm k}=0$, while the former has two split-Dirac points located at ${\bm k}\neq 0$. This may indeed alter the physics of impurity scattering processes, resulting in the distinct behaviors that we observed in our present study. What is certain is that lower temperatures are needed to fully enter the critical regime in the BHZ mode, something that we defer to future investigations.

\section{Conclusions}

In conclusion, the simulations based on the non-commutative Kubo formula and a recently developed numerical implementation enabled us to converge the transport coefficients at temperatures low enough to enter the quantum critical regime at the Chern-to-normal insulator transition, at least for the Kane-Mele model. When compared with the available experimental facts and our previous simulations for PIT in IQHE, the results on the two strongly disordered Chern insulator models show similarities but also important differences. The similarities include: the existence of a single critical point and the single-parameter scaling behavior; The KM model, for which the full critical regime seemed to be reached by our calculations, displays a scaling exponent consistent with the universally accepted value for the unitary class. Among the dissimilarities were the absence of the Quantized Hall Insulator phase, a universal critical value of $\sigma_{xx}^c\approx \frac{e^2}{h}$ instead of $\sigma_{xx}^c\approx \frac{1}{2}\frac{e^2}{h}$, and the violation of the semi-circle law.

\section*{Acknowledgments}

This work was supported by the U.S. NSF grants DMS-1066045 and DMR-1056168. 


%

\end{document}